\documentclass[conference,page-numbering]{IEEEtran}

\usepackage{amsthm}
\usepackage{algorithm}
\usepackage{algpseudocode}
\usepackage{textcomp}
\usepackage{amsmath}
\usepackage{stmaryrd}
\usepackage{graphicx}
\usepackage{syntax}
\usepackage{fancyhdr}
\usepackage{textcomp}
\usepackage[matrix,arrow,curve]{xy}

\theoremstyle{plain} 

\newcommand{\var}[1]{\underline{\texttt{\textit{#1}}}}

\begin{document}
\xyoption{all}

\title{Unification of Template-Expansion and XML-Validation}

\author{\IEEEauthorblockN{Ren\'{e} Haberland}
  \IEEEauthorblockA{Saint Petersburg State University\\
  Saint Petersburg, Russia}
}

\maketitle
\thispagestyle{plain}
\pagestyle{plain}
\pagenumbering{gobble}


\IEEEpeerreviewmaketitle

\textit{Recommended by Professor Bratchikov I.L., 7th April 2008.}
\section*{Extended Abstract}
The processing of XML documents often includes creation and validation. These two
operations are typically performed in two different nodes within a computer network
that do not correlate with each other.

The process of creation is also called \textit{instantiation of a template} and can be described
by filling a template with data from external repositories. Initial access to arbitrary
sources can be formulated as an expression of certain \textit{command languages} like XPath.
Filling means copying invariant element nodes to the target document and unfolding
variable parts from a given template. \textit{Validation} is a descision problem returning true if
a given XML document satisfies a \textit{schema} and false otherwise. The main subject is to
find a language that unions the template expansion and the validation.\\

Schemata and stylesheets, both describe the same XML documents but often they are
not similar. The more precise a schema describes a template the more the schema becomes
complex and confusing. If access to external repositories is described explicitly in
a template language, then the corresponding schema becomes large. Generally template
languages, in particular XSLT, have "unfertile" tags that have no corresponding meaning
in schema languages. Moreover, different schema languages can be characterised
by corresponding schema grammars that can have different computational power. Too
powerful schema grammars effect too coarse-grained schemata in comparison with less
powerful schemata and heterogenous communications and architectures. Consequently
a coarse-grained schema can validate successfully in spite of an impossible validation.
Beside XSLT other template languages vary just in the manner of how they produce
XML. The consequence is to examine an abstraction of the generation process.

Generating automatically a schema from a XML instance document is not a real solution 
because two different languages remain. The main advantage of obtaining rapidly
exploitable schemata is complicated by parametrisation, order, cardinalities and configuration
of included constraints. The idea of static validation is to check the correct
instantiation instead of validating it later on. The specification is often done separately,
so the process consists of one document more than in the "brute-force" approach.

Lingual unification includes syntactical and semantical unification. The same command
tags get different interpretations so that the total amount of documents to be
managed can be reduced in this way. Obviously, tags with such claims require the reduction
of a template language to a real subset. For this very reason it is important to
analyse possible restrictions of the expressiveness.\\

The instantiation takes some template and some external repositories of arbitrary
structure and meaning and creates a XML document. Because we assume that the
template can also be used as a schema we claim those templates are in XML as well,
thus the parts of a template represent semi-structured data. The repository is abstract
and can either be semi-structured, ontological or of any other type. The result is either
an unparsed string or a well-formed element node in case of fulfilling a query during
execution. Staging tags means to ignore the semantic of the tag until its turn has been
achieved. It can also be stated that instantiation is deterministic, thus no two or more
desicions occur while slot filling. Taking into account that queries to repositories can
return manifold results, the restriction to choose the "correct" is done by the concrete
access functions for the corresponding underlying data model.

At validating a XML document the most significant objective is to find out whether
a given document satisfies specific constraints or not. The goal is not to reconstruct the
original document and to check, whether given repositories are correct. Since there exists
a homomorphism between hedge elements, it can be essential if the command language
is taken into account to find a shortened validation plan or using a width-based strategy.
The morphic conditions hold, if and only if referential transparency is guaranteed.

If a given schema grammar is real regular, then the language formed by this grammar
is enclosed under union, intersection and set minus. This condition is sufficient and
such a schema language is extensible for existing tags. In both cases, instantiation
and validation, the corresponding documents are traversed top-down. Templates fulfil
transformations and for that reason only need cycles and selections. To base a validation
only on a \textit{minimal-knowledge}-strategy it is suitable further to restrict recursive calls such
that only unparameterised definitions and calls of subprograms are allowed. As a result
we get a few weak operations that can be used for document processing but not for
algorithmisation.

A schema can sufficiently be well formed by a regular grammar. Continuing this idea,
the model should be explicitely regular to avoid unranked trees and even so by further
normalisation and canonisation can simplify the validation at all.\\

There have been compared three standard schema languages DTD, W3C XSD, RelaxNG and the experimental kernel language for instantiation and validation XTL. For
the expression power XTL and RelaxNG win. Syntactically XTL has the smallest set
of tags but together with RelaxNG they are less complicated than XSD’s definition.
Semantically XSD has the badest result, not only for the overwhelming set of tags that
have no corresponding accordance in a template language. Marks for syntax and semantic
have been made by different criteria and weighted by average demands to schema languages.

\section{Introduction}
The processing of XML documents often includes creation
and validation. These two operations are typically performed in two different nodes within
a computer network that do not correlate with each other. The reasons are specific problems
of distributed systems such as availability, distribution of computing powers, protection aspects
and heterogeneous technologies.
Even if a generated XML-document was successfully created and sent the validation by
an independent node keeps essential.

The process of creation $\Omega$ is also called \textit{instantiation of a template} \cite{Hab:2007} and can be described by filling a template with data from external repositories. Initial access to arbitrary sources can be formulated as an expression of certain \textit{command languages} like XPath.
Filling means copying invariant element nodes to the target document and unfolding variable parts from a given template \cite{Par:2004}. Template languages are XSLT, JSP, ASP, Prolog with extensions and languages using templates as input.
\textit{Validation} $\Theta$ is a descision problem returning true if a given XML document satisfies a \textit{schema} and false otherwise. The schema can be expressed for example in DTD, W3C XSD or RelaxNG. By adding an error report on negative results, validation can be considered as an analysis.

The main aim is to find a language that unions the template expansion and the validation of XML-documents.

\section{Importance}
\subsection*{"Brute-force".} The result of instantiation of a XSLT-stylesheet is not
mandatory XML. Schemata and stylesheets both describe the same XML-documents but often they are not similar. The more precise a schema describes a template the more the schema becomes complex and confusing. If access to external repositories is described explicitly in a template language, then the corresponding schema becomes large. Generally template languages, in particular XSLT, have "unfertile" tags that have no corresponding meaning in schema languages. Moreover, different schema languages can be characterised by corresponding schema grammars that can have different computational power. Too powerful schema grammars effect too
coarse-grained schemata in comparison with less powerful schemata and heterogenous communications and architectures. Consequently a coarse-grained schema can validate successfully in spite of an impossible validation. Beside XSLT other template languages vary just in the manner of how they produce XML. The consequence is to examine an abstraction of the generation process. 

\subsection*{Validation by Instance.} Generating automatically a schema from a XML instance document is not a real solution because two different languages remain. The main advantage of obtaining rapidly exploitable schemata is complicated by parametrisation, order, cardinalities and configuration of included constraints \cite{Moertel}.

\subsection*{Static Validation.} Idea of static validation is to check the correct instantiation
instead of validating later on. The specification is often done separatly, so the processing consists
of one document more than in the "brute-force" approach. But this approach inspires us to
choose such a unified language that is self explanatory \cite{Brics}.

\subsection*{Unified approach.} Lingual unification includes syntactical and semantical unification. The same command tags get different interpretations so that the total amount of
documents to be managed can be reduced in this way. Obviously tags with such claims require the reduction
of a template language to a real subset. For this very reason it is important to analyse possible restrictions of the expressiveness.

After unification template expansion and validation are still kept seperated.
Recalls from the receiver's side can be cancelled, hence every node can perform those operations
by itself.
Since access to external data repositories is obfuscated from the template language the command language will be abstracted.

The unification is supported by outputting only well-formed XML-documents and accessing to external data only by well-typed access functions.

\section{Threads}
In this article the following threads have been analysed:

\begin{itemize}
 \item Which formal properties belong to template expansion and valida\-tion in general and for a
concrete minimalistic unified language?
 \item Which language entails the most syntactical elements, that have an appropriate semantic for
$\Omega$ and $\Theta$?
  \begin{itemize}
    \item What is the aim of $\Omega$ - performing algorithms or document processing?
    \item How do proposed constraints \cite{Cza:1999},\cite{Par:2004} effect $\Theta$?
    \item What happens if $\Omega$ respects command language and the total amount of cycles is determined before? -- This question is not mentioned in this article.
    \item While $\Omega$ the current result set is given to child nodes. That is why expressions in command language are always related to the previous result set. How do we handle those sets in $\Theta$?
  \end{itemize}
 \item How to formalise both processes so that they are similiar as much as possible (taking the extensibility of the language into account)?
 \item Which models are appropriate to describe those processes?
 \item How expressive are languages based on the defined models and following extensions?
 \item How useful is a minimalistic $\Omega$-language in comparison to other common $\Omega$-languages? -- What does the separation of command and template languages effect?
\end{itemize}

\section{Properties of $\Omega$,$\Theta$}
The instantiation $\Omega$ takes some template and some external repositories of arbitrary structure and meaning and creates a XML-document. Because we assume that the template can also be used as schema we claim that templates are in XML as well, thus the parts of a template represent semi-structured data. So the type signature of $\Omega$ is
\texttt{XML$\rightarrow$a$\rightarrow$XML}. Type \texttt{a} is describing a polymorphic type.
That can be either semi-structured, ontological or any other. The result is either an unparsed string or a well-formed element node in case of satisfaction during execution of a query. According to the type, $\Omega$ is an endomorph mapping. Although the expansion of a variable tag can be staged and the mapping type can be preserved. Staging tags means in $\Theta$ to ignore
the semantic of the tag until its turn has been achieved. Since command tags will not be considered while execution of $\Omega$ and $\Theta$, the equation \texttt{$\Omega(\Omega(t,a),a) = \Omega(t,a)$} holds exactly for some template \texttt{t} and some repository \texttt{a}. It can also be stated that $\Omega$ is deterministic, thus no two or more desicions occur while slot filling. Taking into account that queries to repositories
can return manifold results, the restriction to choose the "correct" is done by the concrete
access functions for the corresponding underlying data model.

At validating a XML document the most significant objective is to find out whether a given document satisfies specific constraints or not. The goal is not to reconstruct the original and to check, whether given
repositories are correct. So the signature of $\Theta$ is \texttt{XML$\rightarrow$XML$\rightarrow$Bool}. Because a hedge $x \cdot y$ \cite{Mur:1995} can be validated sequentially, $\theta(x \cdot y) = \theta(x) \wedge \theta(y)$ holds. This homomorphism can be essential if the command language is taken into account to find a shortened validation plan or when a width-based strategy is used. The homomorphic condition
holds, if and only if referential transparency is guaranteed.

If the schema grammar is regular, then the language generated by this grammar is enclosed under union, intersection and set minus. This condition is sufficient \cite{Mur:2001} and so
such a schema language is extensible for existing tags.
Both, $\Omega$ and $\Theta$, are traversing their documents top-down.

\section{Conditions on the models}
Templates fullfill transformations and for that reason \cite{Cza:1999} only need cycles and selections. Another idea \cite{Par:2004} is to restrict allowed operations to attribute references, conditions, recursive calls and conditional inclusion. To base a validation
only on a "minimal-knowledge"-strategy it is suitable further to restrict recursive calls
such that only unparameterised definitions and calls of subprograms are allowed. As a result we get a few weak operations that can be used for document processing but not for algorithmization.

\section{Modelling}
\label{sect:modelling}
As described in \cite{Chi:2000} a schema can be sufficiently well-formed by a regular grammar.
Continuing this approach, the model should be explicitely regular. In detail regular operators have been modelled as ordered binary descision diagrams to break down to unranked trees which need less cases to analyse due to the list of definitions.
\begin{verbatim}
    data Reg = MacroR String |
               AttrR String String |
               TextR String |
               IncludeR String |
               ElR String
                [(String, String)]  Reg |
               TxtR String |
               Epsilon |
               Or Reg Reg |
               Then Reg Reg |
               Star Reg
\end{verbatim}

The order of the descision diagram is right associative and left to right. A
hedge \var{n0}$\cdot$\var{n1}$\cdot$\var{n2}, for example, becomes
\texttt{Then \var{n0} Then \var{n1} Then \var{n2} Epsilon}.

The next step in simplifying validation is normalisation. So, many illegal combinations
can be excluded and legal combinations get an unique canonised form.

Consequently, $\Theta$ can be described by substitution rules. As $\Theta$ works stepwise without any additional correction without dependent processes and exclusivly schema-oriented,
validation equals graph-matching \cite{Rahm:01}. To validate \texttt{IncludeR} it is necessary to
check all occuring element nodes in the given XML document by a disjunctive clause.

\section{Implementation}
\begin{figure}
\begin{center}
\begin{xy}
 (0,0)   = "x1" *\cir<2pt>{} *+!R{\varepsilon},
 (0,-4)  = "x2" *\cir<2pt>{} *+!R{Then},
 (0,-12) = "x3" *\cir<2pt>{} *+!R{ElR},
 (0,-16) = "x4" *\cir<2pt>{} *+!R{TxtR},
 (-4,-8)  = "ellipse1" *\xycircle<30pt,40pt>{},
 (40,8)  = "y1" *\cir<2pt>{} *+!L{\varepsilon},
 (40,4)  = "y2" *\cir<2pt>{} *+!L{Then},
 (40,0)  = "y3" *\cir<2pt>{} *+!L{Star},
 (40,-4) = "y4" *\cir<2pt>{} *+!L{Or},
 (40,-12) = "y5" *\cir<2pt>{} *+!L{ElR},
 (40,-16) = "y6" *\cir<2pt>{} *+!L{TxtR},
 (40,-20) = "y7" *\cir<2pt>{} *+!L{AttR},
 (40,-24) = "y8" *\cir<2pt>{} *+!L{TextR},
 (45,-8)  = "ellipse2" *\xycircle<30pt,65pt>{},
 {\ar "x1";"y1"},
 {\ar "x1";"y2"},
 {\ar "x1";"y3"},
 {\ar "x1";"y4"},
 {\ar "x1";"y5"},
 {\ar "x1";"y6"},
 {\ar "x1";"y7"},
 {\ar "x1";"y8"},
 {\ar "x2";"y1"},
 {\ar "x2";"y2"},
 {\ar "x2";"y3"},
 {\ar "x2";"y4"},
 {\ar "x2";"y5"},
 {\ar "x2";"y6"},
 {\ar "x2";"y7"},
 {\ar "x2";"y8"},
 {\ar "x3";"y1"},
 {\ar "x3";"y2"},
 {\ar "x3";"y3"},
 {\ar "x3";"y4"},
 {\ar "x3";"y5"},
 {\ar "x3";"y6"},
 {\ar "x3";"y7"},
 {\ar "x3";"y8"},
 {\ar "x4";"y1"},
 {\ar "x4";"y2"},
 {\ar "x4";"y3"},
 {\ar "x4";"y4"},
 {\ar "x4";"y5"},
 {\ar "x4";"y6"},
 {\ar "x4";"y7"},
 {\ar "x4";"y8"}
\end{xy}
\caption{Generation of validation rules}
\label{datamodells}
\end{center}
\end{figure}
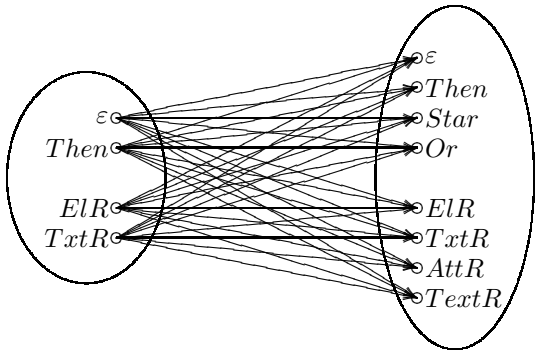

The realisation of nondeterministic transitions in $\Theta$ is made by a nondeterministic splitter that only looks up one element before. In figure \ref{datamodells} the bipartite
graph-matcher is represented. To validate for instance a hedge from a schema against a XML document it is necesarry to refine the \texttt{Then-Then}-case.

\section{Comparison}
There were four schema languages compared: DTD, W3C XSD, RelaxNG and the XML-Template Language
"XTL" based on the proposed model (see \label{sect:modelling}). 

For the expression power the inequality DTD$<$XSD$<$RelaxNG$\leq$XTL holds.
Syntactically XTL has the smallest set of tags but together with RelaxNG they are less
complicated than XSD's definition. Semantically XSD has the badest result, not only for the
overwhelming set of tags that have no corresponding accordance in a template language \cite{Hab:2007}.
Marks for syntax and semantic have been made by different different criteria and weighted
by average demands to schema languages.



\renewcommand{\refname}{References}

\end{document}